\documentclass[12pt]{iopart}

\begin{document}

\title[The influence of interaction between quasiparticles on parametric resonance\ldots]{The influence of interaction between quasiparticles on parametric resonance in 
Bose-Einstein condensates}

\author{Pawe\l{} Zi\'n and Maciej Pylak}
\address{National Centre for Nuclear Research \\
 Ho\.za 69, 00-681 Warsaw, Poland}
\ead{Pawel.Zin@ncbj.gov.pl}

\vspace{10pt}
\begin{indented}
\item[]\today
\end{indented}

\begin{abstract}
We analyse a uniform  weakly interacting bosonic gas
undergoing a periodic oscillation of the interaction constant.
This, within the Bogoliubov approximation, leads to the creation of 
atomic pairs with well defined opposite velocities.
We show how the interaction between quasiparticles, omitted in the Bogoliubov
approximation, significantly changes the atom pair creation process and the properties of the scattered atoms.
\end{abstract}

\pacs{03.75.Gg, 42.50.Dv, 03.75.Kk}
%
\vspace{2pc}
\noindent{\it Keywords}: entanglement, squeezed states, ultracold gases
%

\submitto{\JPB}
%
%
%

\newcommand{\x}{{\bf r}}
\newcommand{\K}{{\bf k}}
\newcommand{\dk}{{\bf \Delta k}}
\newcommand{\KK}{{\bf K}}
\newcommand{\X}{{\bf R}}

\newcommand{\B}[1]{\mathbf{#1}} 
\newcommand{\f}[1]{\textrm{#1}} 

\newcommand{\half}{{\frac{1}{2}}}

\section{Introduction}
\label{Int}

The generation of non-classical states in atomic ensembles is a rapidly developing 
 direction in trapped ion and cold neutral atomic physics \cite{przeglad}.
Such states can be used to increase the sensitivity of precision measurements
beyond the standard classical limit \cite{przeglad2}.
A 100 times decrease of   measurement noise beyond the classical limit 
was recently reported  in cold thermal atoms \cite{Kasevich}.
One of the possible states that are particle entangled, and can be used to 
increase the sensitivity of precision measurements is a co called twin-Fock state $|n,n\rangle$ \cite{limit1,limit2}.
Such a state can be created in experiments generating atomic pairs with well defined momenta
in quasi-one dimensional systems.
This was done by modulation of the atomic interaction parameter
\cite{Paryz1}, modulation instability present in a one dimensional lattice \cite{Paryz2}
or else by the decay of an excited state \cite{Wieden}.
The theoretical analysis for these situations, 
was performed using the Bogoliubov approximation \cite{Wasak,instability,Casimir}.
In this case, the Hamiltonian is quadratic in field operators. 
It has the term responsible for creation of atomic pairs but neglects the terms of higher order
in field operators, which describe the interaction between quasiparticles.
As the process of pair creation starts, the atoms, according to the Bogoliubov description,
are created in pairs with well defined momenta.
Therefore, one expects number squeezing to take place, which is the clear signature of entanglement  \cite{jan}. 
In two of the experiments mentioned   number squeezing was observed \cite{Paryz2,Wieden}.
However it was not seen in the experiment described in \cite{Paryz1}.
Moreover, pair production in \cite{Paryz1} was much smaller than predicted by
the Bogoliubov theory.
This suggests that the interaction between quasiparticles, neglected in the Bogoliubov approximation,
significantly influences the pair production.
Analysis of this influence in the case of modulation of the atomic interaction parameter, as performed in \cite{Paryz1}, 
is the main goal of the present effort.
Such an analysis was performed using a phenomenological model of weak dissipation \cite{busch}.
Here it is achieved using the true microscopic Hamiltonian. 
Let us mention that trapped bosonic gas undergoing periodic oscillation of interaction parameter 
was analysed in the context of Faraday waves \cite{faraday}

The plan of the paper is as follows.
In Section \ref{Sec1}, we analyse the system with time modulated interaction 
within the Bogoliubov method. 
In Sections \ref{s2s} and \ref{SPar} we take the interaction of 
quasiparticles into account. This is done via the Keldysh formalism. 
In Section \ref{s2s} we analyse the system without modulation of the interaction constant. 
Next we introduce an approximation to the self energy functions.
In Section \ref{SPar} we analyse the system with time modulation of the interaction constant. 
Within the approximation of the self energy function, we find an analytical solution
of the Dyson equation. As a result we obtain analytic formulas for the one and two body properties of the system.
A short introduction to the Keldysh formalism, together with lengthy calculations, are moved to the Appendices.

\section{Description of the system. The Bogoliubov method.}\label{Sec1}

The system under consideration is a weakly interacting bosonic gas in a 
three dimensional box with periodic boundary conditions. The Hamiltonian of the system is given by
\begin{eqnarray*}
 \hat{H} = \int \mbox{d} \x \, \frac{\hbar^2}{2m} \nabla \hat{\psi}^\dagger(\x) \nabla \hat{\psi}(\x)
 + \frac{g(t)}{2} \hat{\psi}^\dagger(\x) \hat{\psi}^\dagger(\x)\hat{\psi}(\x) \hat{\psi}(\x),
\end{eqnarray*}
where $\hat \psi(\x)$ is the bosonic field operator and $g(t)$ denotes the interaction constant.
The evolution of the field operator in the Heisenberg picture \cite{Fetter} is then equal to
\begin{equation}\label{psievo}
i \hbar \partial_t \hat \psi(\x,t) = - \frac{\hbar^2}{2m} \nabla^2 \hat \psi(\x,t)+
 g(t)  \hat \psi^\dagger(\x,t) \hat \psi(\x,t) \hat \psi(\x,t).
\end{equation}
The intention of the present work is to analyse the system with a periodic variation 
of the interaction i.e.
\begin{eqnarray*}
 g(t)= g(1+\epsilon \cos(2\omega t) \theta(t)) 
\end{eqnarray*}
where $\epsilon \ll 1$ and $\theta(t)$ is the unit step function.
Let us first do this within the Bogoliubov method. 
Then the field operator is represented as 
\begin{equation}\label{psiform}
 \hat \psi(\x,t) \simeq 
 e^{-i \frac{\mu}{\hbar}t} \left(\sqrt{\frac{N_0}{V}} + \frac{1}{\sqrt{V}}\sum_{\K \neq 0} e^{i\K\x} 
 (u_k \hat b_\K - v_k \hat b_{-\K}^\dagger) \right)
\end{equation}
where $N_0$ is the mean population of the $\K=0$ mode and $\mu$ is the chemical potential.
For temperatures low enough $\mu \simeq n_0 g$ where $n_0 = \frac{N_0}{V}$.
The $\hat b_\K$ are quasiparticle annihilation operators with coefficients
\begin{equation} \label{ukvkwzor}
 u_k = \frac{1}{\sqrt{2}} \sqrt{\frac{E_k + n_0g}{\hbar \omega_\K} +1 }
 \ \ \ \ \ \  
 v_k = \frac{1}{\sqrt{2}} \sqrt{\frac{E_k + n_0g}{\hbar \omega_\K} -1 },
\end{equation}
where $E_k = \frac{\hbar^2 k^2}{2m}$  and 
\begin{eqnarray*}
 \hbar \omega_\K = \sqrt{E_k \left(E_k + 2 n_0g \right)  }
\end{eqnarray*}
is the Bogoliubov energy spectrum.
The Bogoliubov approximation consists of substitution the form (\ref{psiform})
into the equation of motion (\ref{psievo}) and leaving only linear terms in the quasiparticle operators.
Upon doing so, one arrives at
\begin{equation}\label{bdot}
 i \hbar \frac{\mbox{d}\hat b_\K}{\mbox{d} t} = 
 \hbar \omega_k \hat b_\K + 2 \hbar \delta  \cos(2 \omega t)   (\hat b_\K + \hat b^\dagger_{-\K})  
\end{equation}
where
\begin{equation} \label{Edelta}
 \delta = (u_k-v_k)^2 \frac{n_0g \epsilon}{2 \hbar}.
\end{equation}

The equation of motion of the quasiparticle annihilation operator $\hat b_\K$ given by
equation (\ref{bdot}) is linear and can be solved analytically by means of Mathieu functions.
However, if $|\omega - \omega_k| \ll \omega $ and $\delta \ll \omega $ 
we can use the rotating wave approximation in equation (\ref{bdot}). We obtain
\begin{equation}\label{bk}
 i \hbar \frac{\mbox{d}\hat b_\K}{\mbox{d} t} = \hbar \omega_k \hat b_\K + \hbar \delta    
 e^{-2i\omega t} \hat b^\dagger_{-\K}. 
\end{equation}
The solution is
\begin{equation}\label{bt}
\hat b_\K(t)  = \left(\cosh \Omega t + i \frac{\Delta}{\Omega} \sinh \Omega t \right) 
 e^{-i\omega t}\hat b_\K(0)-i \frac{\delta}{\Omega} \sinh \Omega t \,  e^{-i\omega t} \hat b_{-\K}^\dagger(0)
\end{equation}
where $\Delta =  \omega - \omega_k$, $\Omega = \sqrt{\delta^2 - \Delta^2}$.
The evolution of the  $\hat b_\K$ operator shows amplified solutions for $\delta > \Delta$ with resonance condition  $\omega = \omega_\K$. 
This can be seen qualitatively from the Hamiltonian of the system, 
which in the rotating wave approximation takes the form
\begin{equation}\label{H0}
\hat H_0 = \sum_{\K \neq 0}  \hbar \omega_\K \hat b_\K^\dagger \hat b_\K
 + \frac{\hbar \delta}{2} \left( \hat b_\K \hat b_{-\K} e^{2i\omega t} + h.c.  \right).
\end{equation}
If we divide the above Hamiltonian into two parts $\hat H_0 = \hat H_1 + \hat H_2$
where $\hat H_1 = \sum_\K  \hbar \omega_\K \hat b_\K^\dagger \hat b_\K$, then in the interaction picture
the second part of the Hamiltonian takes the form 
\begin{eqnarray*}
\hat H_2(t) &=& e^{i \hat H_1 t/\hbar}  \hat H_2  e^{-i \hat H_1 t/\hbar}=\frac{\hbar \delta}{2} \left( \hat b_\K \hat b_{-\K} e^{2i(\omega - \omega_\K) t} 
 + \hat b_\K^\dagger \hat b_{-\K}^\dagger e^{-2i(\omega - \omega_\K) t}  \right).
\end{eqnarray*}
We clearly see that if $\omega-\omega_\K =0$ then the $\hat H_2(t)$ Hamiltonian leads
to creation of $\K$, $-\K$ pairs of quasiparticles. 

Let us analyse the properties of the modes $\K$ satisfying the resonance condition
$\omega_\K = \omega$. 
We assume that initially (at $t=0$) the system is in a quasiparticle thermal state.
Then the quasiparticle  population is  
\begin{eqnarray} \label{quasipop} 
n_\K(t) =\langle \hat b_\K^\dagger(t) \hat b_\K(t) \rangle  = 
\cosh(2\delta t) n_\K + \sinh^2(\delta t)
\end{eqnarray}
where 
\begin{eqnarray} \label{nk0}
 n_\K =\frac{1}{e^{\beta \omega_\K} -1} =  n_\K(0) =\langle \hat b_\K^\dagger(0) \hat b_\K(0) \rangle.
\end{eqnarray}
In the above we used the fact that $n_\K = n_{-\K}$.
We observe exponential growth of the population for $\delta t \gg 1$.
The particle properties are derived using the connection
\begin{equation}\label{pq2}
 \hat a_\K(t) = e^{-i\frac{\mu}{\hbar} t } \left( u_k \hat b_\K(t) - v_k \hat b_{-\K}^\dagger(t)\right).
\end{equation}
The particle population is
\begin{eqnarray} \label{npt}
n_{p,\K}(t) = \langle \hat a_\K^\dagger(t) \hat a_\K(t) \rangle = (u_k^2 + v_k^2) n_\K(t) + v_k^2.
\end{eqnarray}
As it is directly connected to the quasiparticle population, it also grows in time.
Another important property is the number squeezing parameter defined as
\begin{equation}\label{NS}
s(t) = \frac{ \langle \left( \hat a_\K^\dagger \hat a_\K - \hat a_{-\K}^\dagger \hat a_{-\K} \right)^2 \rangle }{ 
 \langle \hat a_\K^\dagger \hat a_\K \rangle + \langle \hat a_{-\K}^\dagger \hat a_{-\K}  \rangle  }
 = \frac{\langle \Delta \hat{n}_{p,\K}^2(t) \rangle}{n_{p,\K}(t) + n_{p,-\K}(t)}.
\end{equation}
If the above parameter is smaller than unity, the state of the system is called number squeezed.
In \cite{jan} it is shown that if 
\begin{equation}\label{jana}
G^{(2)}_{\K,-\K} > \sqrt{ G^{(2)}_{\K,\K} G^{(2)}_{-\K,-\K} }
\end{equation}
where $ G^{(2)}_{i,j} = \langle \hat a_i^\dagger \hat a_j^\dagger \hat a_j \hat a_i \rangle$
then the state of the system is particle entangled. 
In our case   $n_{p,\K}(t) = n_{p,-\K}(t)$, $G^{(2)}_{\K,\K} = G^{(2)}_{-\K,-\K}$,
and we can rewrite the number squeezing parameter as
\begin{eqnarray*}
s = 1 + \frac{  G^{(2)}_{\K,\K} - G^{(2)}_{\K,-\K}}{n_{p,\K}(t) }. 
\end{eqnarray*}
Thus $s < 1$ implies $ G^{(2)}_{\K,-\K} > G^{(2)}_{\K,\K} = \sqrt{ G^{(2)}_{\K,\K} G^{(2)}_{-\K,-\K}} $
which, according to (\ref{jana}) implies that the state is particle entangled. 
A fast computation using (\ref{pq2}) gives:
\begin{eqnarray}  \nonumber
\Delta \hat{n}_{p,\K}(t) &=&  \hat a_\K^\dagger(t) \hat a_\K(t) - \hat a_{-\K}^\dagger(t) \hat a_{-\K}(t)
\\  \label{qp10}
&=& \hat b_\K^\dagger(t) \hat b_\K(t) - \hat b_{-\K}^\dagger(t) \hat b_{-\K}(t) = \Delta \hat{n}_{\K}(t).
\end{eqnarray}
Additionally, from equation (\ref{bt}) we have
\begin{eqnarray*}
 \Delta \hat{n}_{\K}(t) =   \Delta \hat{n}_{\K}(0),
 \end{eqnarray*}
 which gives 
\begin{equation}\label{sq}
\langle \Delta \hat{n}^2_{p,\K}(t) \rangle = \langle \Delta \hat{n}^2_{p,\K}(0) \rangle. 
\end{equation}
Due to the above equation, the number squeezing decreases in time since the population $n_{p,\K}(t)$ increases. 
Whatever the initial value  of $ \langle (\Delta \hat{n}_{p,\K}(0))^2 \rangle $, $s(t)$ would be below unity after 
some time. The state of the system would be particle entangled.

We additionally mention that the process analysed  can be called parametric amplification
since we obtain amplification of the modes as a result
of a periodic modulation of the interaction constant which is one of the parameters of the system.

In the Bogoliubov method the interaction between quasiparticles is neglected.
In the following sections we shall investigate the impact of such interaction 
on the amplification process.

\section{Interaction between quasiparticles}\label{s2s}

The interaction Hamiltonian in the lowest order in quasiparticle operators takes the form: 
\begin{equation}\label{Hint}
\hat H_{int} = \frac{1}{\sqrt{V}} \sum_{\K_1,\K_2,\K_3}
\delta_{\K_1=\K_2+\K_3} U(\K_1,\K_2,\K_3) \hat b_{\K_1}^\dagger \hat b_{\K_2} \hat b_{\K_3} + h.c.
\end{equation}
where $U(\K_1,\K_2,\K_3)$ is derived in \cite{Castin}:
\begin{eqnarray*}
U(\K_1,\K_2,\K_3) =  g\sqrt{n_0} \left( u_{k_1} (u_{k_2}-v_{k_2})u_{k_3} -  (u_{k_1}-v_{k_1})v_{k_2}u_{k_3} 
+ v_{k_1} v_{k_2}(u_{k_3}-v_{k_3}) \right)
\end{eqnarray*}
where $u_k$ and $v_k$ are given by (\ref{ukvkwzor}).
The other terms omitted  in the Hamiltonian are of higher order in quasiparticle operators and for temperatures low enough  
they should not influence the dynamics. 
The above Hamiltonian is derived taking $g(t) = g$ - neglecting the part $g\epsilon \cos(2 \omega t) $. 
The omitted part adds additional interaction but since
it is multiplied by $\epsilon \ll 1$ its contribution is much smaller than given by the Hamiltonian (\ref{Hint}) and we neglect it.
Now, the Hamiltonian of the system is $\hat H = \hat H_0 + \hat H_{int}$ where $\hat H_0$ is given by (\ref{H0}).
To find the properties of the system governed by such Hamiltonian we 
use standard quantum field theory methods, namely the Keldysh formalism \cite{keldysh,kamenev}.
However, in this section we consider the case without parametric amplification 
 i.e. when the noninteracting Hamiltonian is $\hat H_0 = \sum_\K \hbar \omega_\K \hat b^\dagger_\K \hat b_\K $. 
 The case with parametric amplification is considered in the next section.

Now analyse the single particle Green's function.
A short introduction into the Keldysh formalism is presented in  \ref{ap1}.
In the Keldysh formalism we deal with three independent Green's functions - 
retarded, advanced and Keldysh $G^{R,A,K}(\K,t-t')$. 
The Dyson equation for these functions takes the form
\begin{eqnarray}\label{dysonR}
&& G^{R,A} = G_0^{R,A} + G_0^{R,A} \Sigma^{R,A} G^{R,A}
\\ \label{dysonK}
&& G^K = (1+G^R\Sigma^R) G_0^K(1+\Sigma^A G^A) + G^R \Sigma^K G^A
\end{eqnarray}
where in the formula for $G^{R,A}(\K,t-t')$
the notation $G_0^{R,A} \Sigma^{R,A} G^{R,A}$ 
means 
\begin{eqnarray*}
\int_{-\infty}^\infty \mbox{d} t_1 \, \mbox{d}t_2 \, 
G_0^{R,A}(\K,t-t_1) \Sigma^{R,A}(\K,t_1-t_2) G^{R,A}(\K,t_2-t').
\end{eqnarray*}
Now comes the crucial simplification. 
We assume that $ \Sigma^{R,A,K}(\K,t) e^{i\omega_\K t} $ changes on a 
time scale  much smaller than the time in which $G^{R,A,K}(\K,t)e^{i\omega_\K t}$ changes significantly. 
This leads us to approximate the self energy functions by Dirac delta functions:
\begin{equation}\label{sap}
\Sigma^{R,A,K}(\K,t)e^{i\omega_\K t} = \delta (t) \int \mbox{d} \tau \,  \Sigma^{R,A,K}(\K,\tau) e^{i \omega_\K \tau}
\end{equation}
Note the presence of the factor $e^{i\omega_\K t }$. This is due to the fact that
the functions  $ \Sigma^{R,A,K}(\K,t)$ and $G^{R,A,K}(\K,t)$ oscillate with frequency $\omega_\K$
and the period of oscillation is much smaller than the decay time.
We get rid of this oscillation by multiplying $ \Sigma^{R,A,K}(\K,t)$ and $G^{R,A,K}(\K,t)$ by $e^{i\omega_\K t} $.
For $\Sigma^{R,A,K}$ calculated in second order perturbation theory we obtain
\begin{eqnarray}  \nonumber
\Sigma^R(\K,t) e^{i\omega_\K t} &=& (\Delta_k - i \gamma_k) \delta(t) 
\\  \label{sigmyapp}
\Sigma^A(\K,t) e^{i\omega_\K t} &=& (\Delta_k + i \gamma_k)\delta(t) 
\\ \nonumber
\Sigma^K(\K,t) e^{i\omega_\K t} &=& - 2 i \gamma_k(2n_\K+1) \delta(t) 
\end{eqnarray}
where $\gamma_k > 0$ and $\Delta_k$ is a real number.
The derivation of the above formulas is given in  \ref{sss2}.
The noninteracting Green's functions are given by (\ref{G0n}) 
\begin{eqnarray}\nonumber
iG_0^R(\K,t) &=& \theta(t) e^{-i\omega_\K t} 
\\ \label{prop}
iG_0^A(\K,t) &=& -\theta(-t)e^{-i\omega_\K t} 
\\ \nonumber
iG_0^K(\K,t) &=&  (2n_\K+1)e^{-i\omega_\K t} 
\end{eqnarray}
where $n_\K = \langle \hat b_\K^\dagger \hat b_\K \rangle =\frac{1}{\exp (\beta \hbar \omega_\K) -1}$ 
is the thermal mode occupation.
With this form of Green's function together with approximation (\ref{sigmyapp}), 
the Dyson equation (\ref{dysonR}) for the retarded component is
\begin{eqnarray*}
 iG^R(\K,t) e^{i\omega_\K t}  =  \theta(t)  + \int_{-\infty}^t \mbox{d}t_1 \, (\Delta_k - i \gamma_k) 
 G^R(\K,t_1) e^{i\omega_\K t_1}.
\end{eqnarray*} 
It is easy to check that the solution of the above equation is 
\begin{equation}\label{GRsol}
iG^R(\K,t) = e^{-(\gamma_k + i(\Delta_k + \omega_\K )) t} \theta(t) =
e^{-(\gamma_k + i\Delta_k) t} i G^R_0(\K,t).
\end{equation}
In the same way we obtain
\begin{equation}\label{GAsol}
iG^A(\K,t) = -e^{(\gamma_k - i (\Delta_k+\omega_\K)) t} \theta(-t) = 
e^{(\gamma_k - i \Delta_k) t} 
iG^A_0(\K,t) 
\end{equation}
In the case of $G^K$ it is known that for
stationary problems the first component on the right hand side of equation (\ref{dysonK}) vanishes i.e.
\begin{eqnarray*} 
(1+G^R\Sigma^R) G_0^K(1+\Sigma^A G^A) = 0
\end{eqnarray*}
which turns out to be true in the case of approximation 
(\ref{sigmyapp}).
As a result we are left with
\begin{eqnarray*} 
G^K = G^R \Sigma^K G^A.
\end{eqnarray*}
We substitute in above the analytical form of  
$G^{R,A}$ 
given by (\ref{GRsol}) and (\ref{GAsol}) together with 
the form of $\Sigma^K$ given by (\ref{sigmyapp}).
After performing integrals we obtain
\begin{eqnarray} \label{GKsol}
iG^K(\K,t) &=& (2n_\K+1) e^{-\gamma_k |t| - i (\Delta_k + \omega_\K )t} = iG_0^K(\K,t)e^{-\gamma_k |t|-i \Delta_k t}.  
\end{eqnarray}
We clearly see in equations (\ref{GRsol}),(\ref{GAsol}) and
(\ref{GKsol}) that in all the Green's functions, the interaction between quasiparticles 
leads to decay together with a shift of frequency, as expected \cite{Shi}.

Having analyzed the system within the Dirac delta approximation of the self energy function let
us discuss its validity. 
The calculation of $\Sigma^{R,A,K}(\K,t)$ from first principles though possible, is a demanding task 
beyond the scope of this paper. The authors have never seen such calculations for the system considered here.
Thus the direct way of showing that the effective width in $t$ of the  
$ \Sigma^{R,A,K}(\K,t) e^{i\omega_\K t} $ is much smaller than the time on which $G^{R,A,K}(\K,t)e^{i\omega_\K t}$ changes significantly
is in practice inaccessible. 
However within this approximation the quasiparticle decay turns out to be given by the exponential function.
Such a decay indeed takes place in the system and as known to the authors
this was shown in two independent ways \cite{Castin,Shi}.
First with the use of perturbation calculus analogous to the one presented here but performed in the frequency domain.
Defining
\begin{eqnarray*}
 \Sigma^{R,A,K}(\K,\omega) = \int \mbox{d} t \, e^{i \omega t} \Sigma^{R,A,K}(\K,t) 
\end{eqnarray*}
and in the same way $G^{R,A,K}(\K,\omega)$ the Dyson equation (\ref{dysonR})
take the form:
\begin{equation} \label{Gom}
 G^{R,A}(\K,\omega) =  \left(\left(G_0^{R,A}(\K,\omega) \right)^{-1}  - \Sigma^{R,A}(\K,\omega) \right)^{-1}
\end{equation}
We have $G_0^R(\K,\omega) = (\omega - \omega_\K + i 0)^{-1} $ 
and $G_0^A(\K,\omega) = (\omega - \omega_\K - i 0)^{-1} $. As a result
\begin{equation}\label{Gkw}
  G^{R,A}(\K,\omega) = \left( \omega - \omega_\K - \Sigma^{R,A}(\K,\omega) \right)^{-1}.
\end{equation}
The assumption about exponential decay is  equivalent to the assumption that 
$\Sigma^{R,A}(\K,\omega)$ has no dependence on $\omega$ and equals $\Sigma^{R,A}(\K,\omega_\K)$. Then from equation (\ref{sigmaIRA}) we have
 $ \Sigma^R(\K,\omega_\K) = \Delta_k - i \gamma_k$
and $ \Sigma^A(\K,\omega_\K) = \Delta_k + i \gamma_k$. Substituting those
functions into (\ref{Gkw}) we obtain $G^{R,A}(\K,\omega) = G_0^{R,A}(\K,\omega) e^{- \gamma_k |t| } $.
One can however calculate $G^{R,A}(\K,t)$ using  (\ref{Gom}) without assuming
$\Sigma^{R,A}(\K,\omega) \simeq\Sigma^{R,A}(\K,\omega_\K)  $ but taking the true functional dependence
on $\omega$. Such a calculation is quite complex and goes beyond the scope of the paper, 
but it is described in detail in work of Shi and Griffin \cite{Shi}.
Apart from analytical calculations based on perturbation, exponential decay 
was observed in a finite temperature numerical stochastic calculation for a system considered here 
 (three dimensional weakly interacting Bose gas) \cite{Castin}.
As the decay is given by an exponential function and it is within the Dirac delta approximation,
this fact is an indirect justification of the approximation.

We now comment on the problem of quasiparticle decay in one dimensional system.
The formula (\ref{sigmaC} ) for $\Sigma(\K,t_1-t_2)$ given in \ref{sss2} in the frequency domain takes the form 
\begin{eqnarray}\nonumber
&& \Sigma(\K,\omega) \propto
\\ \nonumber
&& \frac{1}{V} \sum_{\K_1} 
\left( 2 U^2(\K,\K_1,\K-\K_1) G_0(\K_1,\omega_{\K_1})G_0(\K-\K_1,\omega_{\K-\K_1}) \delta(\omega-\omega_{\K_1}-\omega_{\K-\K_1}) \right.
\\ \nonumber
&& + \left. 4  U^2(\K,\K_1,\K+\K_1)  G_0(\K+\K_1,\omega_{\K+\K_1})G_0(\K_1,\omega_{\K_1}) 
\delta(\omega+\omega_{\K_1}-\omega_{\K+\K_1})  \right)
\end{eqnarray}
When in the above formula we change the sum into the integral i.e. 
$\frac{1}{V}\sum_{\K_1} \rightarrow \frac{1}{(2\pi)^d} \int \mbox{d} \K_1 $ where $d$ denotes the dimension of the system,
after performing  Dirac delta we arrive for $d=3$ at the integral over two dimensional hyper-surface.
It results in finite result. In the case of one dimensional system $d=1$ we arrive at zero dimensional hyper-surface and
the integral cannot be performed.
This shows that the above presented standard method does not work in the one dimensional case.
The method  to solve such problem is described in \cite{Fetter,gorkov}.
In this method  we have $G_0$ being replaced by true $G$ in the above formula for $\Sigma$.
Then the Dyson equation (\ref{Gom}) becomes nonlinear equation for $G$.
As a result one needs to adopt variational ansatz to solve it.
This was done in \cite{rusek}. As a result the quasiparticle lifetime was found.
However an important question is what is the quasiparticle decay function.
The authors did not found any paper analysing it. 
However by using methods similar to the one used in \cite{lam} we found that the
quasiparticle decay is not given by exponential function \cite{pozniej}.

The reader may be surprised that we did not start from performing the calculation in the frequency domain,
as it is an obvious choice for a time independent Hamiltonian.
The reason is that we want to use the approximation for the parametric amplification case
where the Hamiltonian  depends on time and the use of the frequency domain is rather impractical.

\section{Parametric amplification}\label{SPar}

Let us now turn our attention to parametric amplification. 
Since now the $H_0$ Hamiltonian is given by equation (\ref{H0}), we deal with nonzero
observables $\langle \hat b_\K \hat b_{-\K} \rangle$ and 
$\langle \hat b_\K^\dagger \hat b_{-\K}^\dagger \rangle$.
As a consequence we need to define a new type of Green's function. It turns out that 
it is most convenient to define the matrix Green's function
\cite{Fetter}
\begin{eqnarray} \nonumber
i{\bf G}(\K,t,t') &=&
\left(\begin{array}{cc}
    iG_{11}(\K,t,t')  & iG_{12}(\K,t,t')  \\
    iG_{21}(\K,t,t')  & iG_{22}(\K,t,t') 
   \end{array} \right)
\\
\label{GM}
&=&
\left( \begin{array}{cc}
    \langle T_C [ b_\K(t) b_\K^\dagger(t') ] \rangle &   \langle T_C [ b_\K(t) b_{-\K}(t') ] \rangle 
   \\
   \langle T_C [ b_{-\K}^\dagger(t) b_\K^\dagger(t') ] \rangle    & \langle T_C [ b^\dagger_{-\K}(t) b_{-\K}(t') ] \rangle
   \end{array} \right) 
   \end{eqnarray}
 with the property $G_{22}(-\K,t',t) = G_{11}(\K,t,t')$, together with a matrix self energy function
 \begin{eqnarray} \nonumber
{\bf \Sigma}(\K,t,t') &=&
\left(\begin{array}{cc}
    \Sigma_{11}(\K,t,t') & \Sigma_{12}(\K,t,t') \\
    \Sigma_{21}(\K,t,t') & \Sigma_{22}(\K,t,t')
   \end{array} \right) 
\\ 
\label{SM}
&=&
\left(\begin{array}{cc}
    \Sigma_{11}(\K,t,t') & \Sigma_{12}(\K,t,t') \\
    \Sigma_{21}(\K,t,t') & \Sigma_{11}(-\K,t',t)
   \end{array} \right).
\end{eqnarray}
Please notice that  also $\Sigma_{22}(\K,t,t') = \Sigma_{11}(-\K,t',t) $ 
as in case of the Green's function. 

Further on we perform an approximation to the noninteracting
Green's function $G_0$. These functions appear in the Dyson equation 
which in general has the form $G = G_0 + G_0 \Sigma G$
in two ways. First $G_0$ appears there directly as can be seen.
In the second way it appears in $\Sigma$.
We only approximate  the form of $G_0$ appearing in $\Sigma$ and 
we do it in the following way. 
As can be seen in the solution for $\hat b_\K(t)$ given by (\ref{bt}),
the amplification is present for $|\Delta| < \delta$ - which gives us a region
in $\K$ space which is amplified. Strictly speaking, there is a width of 
$\Delta k$ around $k_0$ for which $\K$ modes satisfying $|k-k_0| < \Delta k$ are amplified. As $|\Delta|$ starts to be larger than $\delta$
the solution (\ref{bt}) tends towards $\hat b_\K(t) = e^{-i\omega_\K t}\hat b_\K(0)$.  Let us now assume that $\Delta k \ll k_0$.
The self energies $\Sigma_{11,12,21,22}$ are introduced as sums over products of the Green's function over all the modes. 
Under this assumption, only a very small part of this sum  
is  contributed by amplified modes.
We therefore perform an approximation and assume that all the modes
while contributing to the self energy functions are 
$\hat b_\K(t) = \hat b_\K(0) e^{-i\omega_\K t}$.
The same evolution was used in the previous section while analysing the case without  parametric amplification. 
We therefore obtain  that the  ${\bf \Sigma}(\K,t,t') $ matrix is determined by  
 $\Sigma(\K,t,t') $ present in the scalar case. Strictly speaking we have $\Sigma_{12}(\K,t)=\Sigma_{21}(\K,t) = 0 $ 
and $\Sigma_{11}(\K,t,t') = \Sigma_{22}(-\K,t',t) = \Sigma(\K,t,t') $ given by equation (\ref{sigmaC}). These facts are discussed in more details in  \ref{KeldyshPar}.

Similar as in the scalar case, we deal with three independent matrix Green's functions ${\bf G}^{R,A,K}$.
The structure of the method is the same as in the scalar case, and the Dyson equation takes the same form
as in (\ref{dysonR}) and (\ref{dysonK}), namely
\begin{eqnarray}\label{dysonRM}
&& {\bf G}^{R,A} = {\bf G}_0^{R,A} + {\bf G}_0^{R,A} {\bf \Sigma}^{R,A} {\bf G}^{R,A}
\\ \label{dysonKM}
&& {\bf G}^K = (1+{\bf G}^R{\bf \Sigma}^R) {\bf G}_0^K(1+{\bf \Sigma}^A {\bf G}^A) + {\bf G}^R {\bf \Sigma}^K {\bf G}^A
\end{eqnarray}
where 
\[{\bf G}^{R,A,K}(\K,t_1,t_2) = 
\left(\begin{array}{cc}
    G_{11}^{R,A,K}(\K,t_1,t_2) & G_{12}^{R,A,K}(\K,t_1,t_2)   \\
    G_{21}^{R,A,K}(\K,t_1,t_2) & G_{11}^{A,R,K}(-\K,t_2,t_1)   
   \end{array} \right) 
 \]
and 
\begin{eqnarray*}
 {\bf \Sigma}^{R,A,K}(\K,t,t') &=&
\left(\begin{array}{cc}
    \Sigma_{11}^{R,A,K}(\K,t,t') & 0 \\
    0 & \Sigma_{22}^{R,A,K}(\K,t',t)
   \end{array} \right)
\\
   &=&
\left(\begin{array}{cc}
    \Sigma^{R,A,K}(\K,t,t') & 0 \\
    0 & \Sigma^{A,R,K}(-\K,t',t)
   \end{array} \right)
 \end{eqnarray*}
where $\Sigma_{11}=\Sigma$.
Proceeding as before, we introduce the assumption about the time scales of $\Sigma^{R,A,K}$.
We arrive at
\begin{eqnarray*}
&&\left(\begin{array}{cc}
    \Sigma_{11}^{R}(\K,t) e^{i\omega_\K t} & 0 \\
    0 & \Sigma_{22}^{R}(\K,t) e^{-i\omega_\K t}
   \end{array} \right)
= \delta(t) 
\left(\begin{array}{cc}
    \Delta_k-i\gamma_k & 0 \\
    0 & \Delta_k + i \gamma_k
   \end{array} \right), 
\\
&& \left(\begin{array}{cc}
    \Sigma_{11}^A(\K,t) e^{i\omega_\K t} & 0 \\
    0 & \Sigma_{22}^A(\K,t) e^{-i\omega_\K t}
   \end{array} \right)
= \delta(t) 
\left(\begin{array}{cc}
    \Delta_k+i\gamma_k & 0 \\
    0 & \Delta_k - i \gamma_k
   \end{array} \right),
\end{eqnarray*}
and
\begin{eqnarray*}
 \left(\begin{array}{cc}
    \Sigma_{11}^K(\K,t) e^{i\omega_\K t} & 0 \\
    0 & \Sigma_{22}^K(\K,t) e^{-i\omega_\K t}
   \end{array} \right)
\delta(t) (-2i) \gamma_k(n_\K+1)
\left(\begin{array}{cc}
    1 & 0 \\
    0 & 1
   \end{array} \right). 
\end{eqnarray*}
The solution of the Dyson equation (\ref{dysonRM}) using the above form of approximate self energies is 
\begin{eqnarray*} \nonumber
i{\bf G}_0^R(\K,\Delta,t,t') = 
\left(\begin{array}{cc}
    A_R(\Delta,t,t')   &  B_R(\Delta,t,t')   \\
   -B_R^*(\Delta,t,t') &  - A_R(\Delta,t',t)   
   \end{array} \right) \theta(t-t') 
\end{eqnarray*}
where the functions $A_R$ and $B_R$ are defined in the \ref{KeldyshPar}. 
Notice that the parametric process is defined by the values of $\delta$
and $\omega$ in the Hamiltonian (\ref{H0}).
The $A_R$ and $B_R$ functions present in the formula for ${\bf G}_0$ 
depend on a single parameter $\Delta= \omega-\omega_\K$.
The Dyson equation (\ref{dysonRM}) takes the form
\begin{eqnarray*} 
&& i{\bf G}^R(\K,t,t') = 
\left(\begin{array}{cc}
    A_R(\Delta,t,t')   &  B_R(\Delta,t,t')   \\
   -B_R^*(\Delta,t,t') &  - A_R(\Delta,t',t)   
   \end{array} \right) \theta(t-t') 
\\ 
&& +  
 \int_{-\infty}^t \mbox{d} t_1 \,
\left(\begin{array}{cc}
    A_R(\Delta,t,t_1)   &  B_R(\Delta,t,t_1)   \\
   -B_R^*(\Delta,t,t_1) &  - A_R(\Delta,t_1,t)   
   \end{array} \right) 
\left(\begin{array}{cc}
    \Delta_k-i\gamma_k & 0 \\
    0 & \Delta_k + i \gamma_k
   \end{array} \right)
{\bf G}^R(\K,t_1,t').
\end{eqnarray*}
The solution is
\begin{eqnarray}\label{GPR}
i{\bf G}^R(\K,t,t') = 
  i{\bf G}^R_0(\K,\Delta-\Delta_k,t,t')e^{-\gamma_k(t-t')}.  
\end{eqnarray}
Let us notice that ${\bf G}_0(\K,\Delta-\Delta_k,t,t')$, being a part of the above solution,
has effective $\Delta$ equal to $\Delta-\Delta_k$.
This is the frequency shift caused by the interaction. 
It was present in the scalar case as well.
In the Bogoliubov method, parametric resonance is obtained 
when $\omega=\omega_\K$ as described in Section \ref{Sec1}.
Here the resonance is for $\Delta - \Delta_k = 0$ which gives $\omega = \omega_\K + \Delta_k$ which gives the shift of frequency
as expected.
Proceeding in the same way, we obtain that
\begin{equation}\label{GPA}
{\bf G}^A(\K,t,t') = {\bf G}^A_0(\K,\Delta-\Delta_k,t,t') e^{-\gamma_k(t'-t)} 
\end{equation}
Having this result, we now choose $\Delta = \Delta_k $. 
It turns out that, as in the scalar case, 
$(1+{\bf G}^R{\bf \Sigma}^R) {\bf G}_0^K(1+{\bf \Sigma}^A {\bf G}^A)=0 $.
As a result the formula for ${\bf G}^K$ takes the form
\begin{eqnarray*}
 {\bf G}^K(\K,t,t') =
  \int \mbox{d}t_1 \, {\bf G}^R(\K,t,t_1){\bf \Sigma}^K(\K,t_1){\bf G}^A(\K,t_1,t')
 \end{eqnarray*}
where the ${\bf G}^{R,A}$ functions are taken for $\Delta=\Delta_k$ and  we used the approximation described in the 
previous Section $ {\bf \Sigma}^K(\K,t_1,t_2)  = {\bf \Sigma}^K(\K,t_1) \delta(t_1-t_2)$.
Using (\ref{GPR}) and (\ref{GPA}) we obtain that for $t \geq t' \geq 0$:
\begin{equation} \label{GPK}
i{\bf G}^K(\K,t,t') = (2n_\K+1)
\left(\begin{array}{cc}
   a  & -i b \\
   i b &  a   
   \end{array} \right) 
\end{equation}
where 
\begin{eqnarray} \nonumber
 a &=&  e^{-\gamma_k(t-t')} 
  \frac{ \gamma_k \left(\gamma_k \cosh(\delta(t-t')) + \delta 
 \sinh(\delta(t-t'))\right)}{ \gamma_k^2 - \delta^2}
\\ \nonumber
& & -e^{-\gamma_k(t+t')} \frac{ \delta \left(\delta \cosh(\delta(t+t')) + \gamma_k
 \sinh(\delta(t+t'))\right)}{ \gamma_k^2 - \delta^2} 
\\ \label{ab}
\\ \nonumber
b &=&  e^{-\gamma_k(t-t')} \frac{ \gamma_k \left( \delta\cosh(\delta(t-t')) + \gamma_k 
 \sinh(\delta(t-t') ) \right) }{ \gamma_k^2 - \delta^2 }
\\ \nonumber
& &
 -   e^{-\gamma_k(t+t')}  
 \frac{\delta  \left(\gamma_k \cosh(\delta (t+t'))  + \delta \sinh(\delta(t+t')) \right)  }{ \gamma_k^2 - \delta^2 }.
 \end{eqnarray}

 We now use the above solution to investigate the properties of the system. 
All below are derived in \ref{nsp}. First, we analyse the quasiparticle population $n_\K(t)$ equal to
\begin{eqnarray}
\nonumber
 2 n_\K(t) +1 &=& iG_{11}^K(\K,t,t) \\
\label{nkt}
&=&
(2n_\K+1)\frac{e^{-2 \gamma_k t} \delta \left(\delta \cosh(2\delta t) +\gamma_k
 \sinh(2\delta t)\right) - \gamma_k^2}{ \delta^2 -\gamma_k^2 } 
\end{eqnarray} 
Looking at the above formula we clearly distinguish two regimes:
\\
1) $\delta > \gamma_k$ where the population grows exponentially in time and 
$\delta t \gg 1$ is equal to
\begin{equation} \label{pop1}
 n_\K(t) \simeq (2n_\K+1) \frac{\delta}{4(\delta-\gamma_k)} e^{2(\delta - \gamma_k)t}.
\end{equation}
2) $\gamma_k > \delta$ where the population reaches its maximum equal to
\begin{eqnarray}\label{nksecond}
 n_\K(\infty)   = \frac{1}{2} \left((2n_\K+1) \frac{\gamma_k^2}{ \gamma_k^2 - \delta^2 } -1 \right)
\end{eqnarray}
The particle population is directly connected with the quasiparticle one described above through relation (\ref{npt}).
Having analysed the population we now turn our attention to the number squeezing parameter given by (\ref{NS}). 
Here we use the lowest order approximation (Wick's theorem)
\begin{eqnarray*}
 \langle \hat b_\K^\dagger \hat b_\K \hat b_\K^\dagger \hat b_\K \rangle &\simeq& n_\K(t)(2 n_\K(t)+1)
 \\
 \langle  \hat b_\K^\dagger \hat b_\K  \hat b_{-\K}^\dagger \hat b_{-\K} \rangle &\simeq& n_\K(t)n_{-\K}(t) 
 + \langle \hat b_\K \hat b_{-\K} \rangle \langle \hat b_\K^\dagger \hat b_{-\K}^\dagger \rangle
\end{eqnarray*}
The terms above Wick are proportional to  $1/V$ and vanish in the thermodynamic limit.
The justification of this fact is given in \ref{Wick}.
In the present paper we do not want to discuss the finite size effects and therefore we neglect this contribution. 
Then the numerator of the number squeezing parameter is
\begin{eqnarray*}
 \langle \Delta n_\K^2(t) \rangle 
&=& 2 n_\K(t)(n_\K(t)+1) - \frac{1}{2} \left|G_{12}^K(\K,t,t)\right|^2
\\
\\
&=& \frac{1}{2} \left( (2n_\K+1)^2 \frac{ 2e^{-2\gamma_k t} \gamma_k\delta \sinh(2\delta t) + e^{-4 \gamma_k t} \delta^2 - \gamma_k^2 }{
\delta^2 -\gamma_k^2 } - 1 \right).
\end{eqnarray*}
Using the above formulas together with (\ref{npt}), (\ref{NS}) and (\ref{qp10}) we find
that in the first regime $\delta > \gamma_k$ for $\delta t \gg 1$:
\begin{eqnarray*}
 s(t) \simeq \frac{2n_\K+1}{u_k^2+v_k^2} \frac{\gamma_k}{\delta+\gamma_k},  
\end{eqnarray*}
while for the second regime $\gamma_k > \delta$ we have
\begin{eqnarray*}
 s(\infty) = \frac{ (2n_\K+1)^2 \frac{\gamma_k^2}{\gamma_k^2 - \delta^2} -1 
 }{2 \left( (2n_\K+1) \frac{\gamma_k^2}{\gamma_k^2 - \delta^2} -1 \right)(u_k^2+v_k^2)+4 v_k^2}.
\end{eqnarray*}
We see that, in both regimes, the number squeezing parameter tends to a non-zero value. It depends on the parameters of the system 
was well as the chosen value of $k$. One must have all the parameters to check whether $s$
is smaller than unity and as a result is particle entangled.
But what is clear is that if $\gamma_k $ is significantly larger than $\delta$
the production process is practically frozen as given by equation (\ref{nksecond}). 
Then the system is practically of no use as a source of 
atom pairs.
But even if $\delta > \gamma_k$ when the atomic pair production is satisfactory,
 the number squeezing parameter 
can still be above unity, making the source useless for increasing measurement sensitivity above the classical limit.
These results modify the one obtained within
the Bogoliubov method, where the number of quasiparticles grows exponentially 
leading eventually to the number squeezing parameter being below unity.

We now comment on the use of the above result in the case of one dimensional system which
is directly connected to the experiment \cite{Paryz1}.
As written in the previous Section the quasiparticle decay function in the one dimensional system is not
given by exponential function.  
Thus, strictly speaking we cannot apply the above results to one dimensional case.
Still the quasiparticle decay function has some width and some average frequency shift.  
Using that we can establish effective $\gamma_k$ and $\Delta_k$.
Than one would expect that the above results are valid i.e. 
the amplification shall be very small if effective $\gamma_k$ is much larger than
$\delta$. 
This would explain very small amplification as compared to that predicted by Bogoliubov method
observed in experiment \cite{Paryz1}.

Finally, we now comment on the connection of the above results with that obtained in \cite{busch}.
There a phenomenological model is used in which the $\gamma_k$ and $\Delta_k$ coefficients appear
as the ones that need to be specified by microscopic theory.
The authors consider the quasiparticle properties of the system, together with the
quasiparticle entanglement criteria.
The authors arrive at the same formulas for the quasiparticle properties as
obtained here. 
However particle and quasiparticle entanglement is not equivalent. 
It can be shown that the quasiparticle entanglement criterion used in \cite{busch}
is stronger than the particle entanglement criterion i.e. $s < 1$ used here.
This means that if the criterion used in \cite{busch} is satisfied we always have $s<1$,
but the fact that $s<1$ does not imply that the criterion is satisfied.

\section{Summary}

We  analysed a uniform system of weakly interacting
bosons undergoing periodic oscillation of an interaction parameter.
We showed, that within the Bogoliubov approximation, this leads to
creation of atom pairs with well defined opposite velocities.
This leads to a number squeezed state particle entangled and useful 
in increasing measurement sensitivity above classical limit. 
We analysed the impact of interaction between quasiparticles, 
neglected in the Bogoliubov approximation, on the atom pair production process.
We showed that this interaction can drastically change the atom creation process.
Strictly speaking, the parametric process is described by a single parameter $\delta$ describing the
strength of the amplification, present in the Bogoliubov method.
Within the approximation, the interaction between quasiparticles 
is described by a quasiparticle decay constant $\gamma_k$ and frequency shift $\Delta_k$ of the quasiparticle energy.

Within the Bogoliubov approximation, the resonance condition 
(when the pair production is the largest) takes the form $\omega = \omega_\K$
where $2\omega$ is the frequency of temporal change of the interaction constant and $\hbar \omega_\K$
is the quasiparticle energy.
The interaction between quasiparticles changes the resonance condition by introducing the frequency shift,
namely $\omega = \omega_\K + \Delta_k$, which in dilute gases is a slight change $|\Delta_k| \ll \omega_\K$ \cite{Shi}.
The crucial change with respect to the Bogoliubov's approximation caused by the interaction between quasiparticles
is given by the decay constant $\gamma_k$. We have identified two regimes of pair production.
First when $\gamma_k > \delta$ the number of pairs increases towards finite limit.
When  $\gamma_k$ is a few times larger than $\delta$ then the increase of pairs is practically zero
- the pairs are not produced making the system useless as a source of atom pairs.
In the second regime when $\delta > \gamma_k$, the pair production is exponential in time 
proportional to $\exp \left( 2 (\delta - \gamma_k) t \right)$. So after some time 
a lot of pairs are produced. 
Additionally we have analyzed the value of the number squeezing parameter.
We found that in both regimes, depending on parameters of the system, it may be below unity.
Than the quantum state is particle entangled an may be 
used in increasing measurement sensitivity above classical limit.

This results are obtained for a three dimensional system. However,  
if used for the one-dimensional case, they provide a possible explanation
of the small amplification observed experimentally in \cite{Paryz1}.

\ack

We acknowledge  discussions with Chris Westbrook, Denis Boiron and Dimitri Gangardt.
P. Z. was supported by the National Science Centre Grant
No. DEC-2011/03/D/ST2/00200.

\appendix

\section{Keldysh formalism} \label{ap1}

\subsection{Introduction into the formalism}
\label{ss1}

An detailed introduction into the Keldysh method can be found in \cite{keldysh,kamenev}.
In the Keldysh formalism we have a contour  from $-\infty$ to $\infty$ and back.
As in the traditional formulation of quantum field theory we introduce a Green's function 
\begin{equation}\label{Green's}
iG(\K,t,t') = \langle T_C[ \hat b_\K(t) \hat b_\K^\dagger(t')] \rangle. 
\end{equation}
The difference with the standard, zero temperature, formulation is that $t$ and $t'$ are now variables on a Keldysh contour with $T_C$ 
being a time ordering operator on that contour.  
It turns out that the Dyson equation holds with  integration  over the Keldysh contour:
\begin{eqnarray} \label{Dyson}
 G(\K,t,t')  &=&  G_0(\K,t,t')+\int_C \mbox{d} t_1 \mbox{d} t_2 \,  G_0(\K,t,t_1) \Sigma(\K,t_1,t_2) G(\K,t_2,t').
\end{eqnarray}
Here $\int_C$ denotes integration over the Keldysh contour and the $G_0$ is the noninteracting Green's function
where the time evolution of the system is only due to $H_0$ Hamiltonian.

A calculation performed on such a contour is impractical and is replaced by real time integrals.
To do that we divide the contour into two parts:  $C_+$  from $-\infty$ to $\infty$,
and $C_-$ going back to $-\infty$. Each quantity present in the 
Dyson equation can have arguments $t$ and $t'$ located on both parts of the contour. 
This gives four possibilities: 
\begin{itemize}
 \item $T$ when $t,t' \in C_+$
 \item $<$ - $t \in C^+$, $t' \in C_-$
 \item $>$ - $t \in C^-$, $t' \in C_+$,
 \item  $\tilde T$ - $t \in C_-$, $t' \in C_-$.
\end{itemize}
As a result, instead of one Green's and self energy function defined on a contour we end up
with four different types of functions $G^{T,<,>,\tilde T}$ and $\Sigma^{T,<,>,\tilde T}$
defined on a real axis.
However it turns out that the property
\begin{eqnarray}\label{rel} 
T + {\tilde T} &=& >+< 
\end{eqnarray}
takes place for $G$, $G_0$ and $\Sigma$ functions (by this we mean, for example $G^T+G^{\tilde T} = G^>+G^<$). 
As a result only three of the above are linearly independent. They are chosen as
\begin{eqnarray} \label{def2}
 R = T - <
\ \ \ \ \ \ 
 A = T - > 
\ \ \ \ \ \ 
 K = >+<. 
\end{eqnarray}
Additionally the {\it retarded} and
{\it advanced} Green's, self energy functions $(G_0,G,\Sigma)^R(\K,t,t')$ and $(G_0,G,\Sigma)^A(\K,t,t')$ are zero for $t'>t$ and $t>t'$ respectively.
The Dyson equation (\ref{Dyson}) now takes the form
\begin{eqnarray} \label{Dyson1}
&& G^{R,A} =G_0^{R,A} + G_0^{R,A} \Sigma^{R,A} G^{R,A} 
\\ \label{Dyson3}
&& G^K = (1+G^R\Sigma^R)G_0^K(1+\Sigma^A G^A) + G^R\Sigma^KG^A. 
\end{eqnarray}

\subsection{Derivation of  $\Sigma^{R,A,K}$}\label{sss2}

Having derived the Dyson equations, we now turn our attention 
to calculating self energies $\Sigma^{R,A,K}$.
In the second order perturbation theory with interaction Hamiltonian given by (\ref{Hint}) we obtain
\begin{eqnarray}\nonumber
&& \Sigma(\K,t_1-t_2) =
\\ \nonumber
&& \frac{i}{V} \sum_{\K_1} 
\left( 2 U^2(\K,\K_1,\K-\K_1) G_0(\K_1,t_1,t_2)G_0(\K-\K_1,t_1,t_2) \right.
\\ \label{sigmaC}
&& + \left. 4 
 U^2(\K,\K_1,\K+\K_1) 
G_0(\K+\K_1,t_1,t_2)G_0(\K_1,t_2,t_1) \right)
\end{eqnarray}
In the above, the component $T,<,>,\tilde T$ is obtained by
taking $t_1$ and $t_2$ on the correct $C_+$ or $C_-$ part.
For example if we want to calculate $\Sigma^T$ then $t_1 \in C_+ $ and $t_2 \in C_+$. 
This implies that the $G_0$ functions  in (\ref{sigmaC}) become $G^T_0$ functions. 
However, in the case of $<$ component $t_1 \in C_+$ and $t_2 \in C_-$ 
which make the $G_0$ functions present in (\ref{sigmaC}) to turn into
\begin{eqnarray*}
G_0(\K_1,t_1,t_2) \rightarrow G_0^<(\K_1,t_1,t_2)
\ \ \ \ \ \ 
G_0(\K_1,t_2,t_1) \rightarrow G_0^>(\K_1,t_2,t_1).
\end{eqnarray*}
Notice that in the second line of the above equation we have the term $G_0^>(\K,t_2,t_1) $ 
with $>$ component due to the fact that time $t_1$ and $t_2$ are inverted in this function.
Having those, we can calculate $\Sigma^R = \Sigma^T - \Sigma^<$
which take the form
\begin{eqnarray}
\nonumber 
\Sigma^R(\K,t_1,t_2) =\frac{i}{V}\sum_{\K_1} \left(
2 U^2(\K,\K_1,\K-\K_1)\right.\\
\nonumber
\left.\left( G_0^T(\K_1,t_1,t_2)G_0^T(\K-\K_1,t_1,t_2) -G_0^<(\K_1,t_1,t_2)G_0^<(\K-\K_1,t_1,t_2)
 \right) \right. \\ 
\nonumber
 + \left.4U^2(\K,\K_1,\K+\K_1)\right.\\
 \left.\left( G_0^T(\K+\K_1,t_1,t_2)G_0^T(\K_1,t_2,t_1)- G_0^<(\K+\K_1,t_1,t_2)G_0^>(\K_1,t_2,t_1) \right)
\right). 
\label{SR}
\end{eqnarray}

Up to now, the Hamiltonian $\hat H_0$ was not specified.
The results of this Appendix are used in Section \ref{s2s}. Therefore we
carry on the calculations with the same noninteracting Hamiltonian as in that section,
 namely $\hat H_0 = \sum_\K \hbar \omega_\K \hat b_\K^\dagger \hat b_\K$.
As now the Hamiltonian $\hat H_0$ does not depend on time, all  quantities
 in the Dyson equation depend on time difference. 
For example, $G_0(\K,t,t') \rightarrow G_0(\K,t-t')$.
To calculate $\Sigma^R$ for the chosen $\hat H_0$ we must first calculate the  Green's functions $G_0$. 
As in the main body of the text, we take the 
thermal state as the initial one.
We obtain
\begin{eqnarray}
\nonumber 
iG_0^T(\K,t-t') = \langle T[ \hat b_\K(t) \hat b_\K^\dagger(t')] \rangle 
= \left((n_\K+1) \theta(t-t') + 
 n_\K \theta(t'-t) \right) e^{-i\omega_\K (t-t')}
\\ \nonumber
iG_0^<(\K,t-t') =\langle  \hat b_\K^\dagger(t') \hat b_\K(t) \rangle 
= n_\K  e^{-i\omega_\K (t-t')}
\\ \label{G0def}
iG_0^>(\K,t-t') = \langle \hat b_\K(t) \hat b_\K^\dagger(t') \rangle =  (n_\K+1)  e^{-i\omega_\K (t-t')}
\\ \nonumber
iG_0^{\tilde T}(\K,t-t') =\langle \tilde T[ \hat b_\K(t) \hat b_\K^\dagger(t')] \rangle 
= \left(n_\K \theta(t-t') + 
 (n_\K+1) \theta(t'-t) \right) e^{-i\omega_\K (t-t')}
\end{eqnarray}
where $n_\K = \langle \hat b_\K^\dagger \hat b_\K \rangle =\frac{1}{\exp (\beta \hbar \omega_\K) -1}$ 
is the thermal mode occupation.
 Now we substitute into (\ref{SR}) the $G_0$ functions given by (\ref{G0def}) to obtain
\begin{eqnarray} \nonumber
\Sigma^R(\K,t) &=& - \theta(t) \frac{i}{V} \sum_{\K_1} 
\left( 2 U^2(\K,\K_1,\K-\K_1) (n_{\K_1}+n_{\K-\K_1}+1)
e^{- i (\omega_{\K_1} + \omega_{\K-\K_1}) t}  
\right.
\\
&+& \left. 4 U^2(\K,\K_1,\K+\K_1) 
(n_{\K_1} - n_{\K+\K_1}) e^{- i (\omega_{\K+\K_1} - \omega_{\K_1})t}  
\right) 
\end{eqnarray}
The above can be rewritten as
\begin{equation}\label{sigmaR}
\Sigma^R(\K,t) = - 2i f(\K,t) e^{-i \omega_\K t}\theta(t)
\end{equation}
where
\begin{eqnarray} \nonumber 
f(\K,t) &=&  \frac{1}{V} \sum_{\K_1} 
\left( U^2(\K,\K_1,\K-\K_1) (n_{\K_1}+n_{\K-\K_1}+1)
e^{i ( \omega_\K -\omega_{\K_1} - \omega_{\K-\K_1}) t}  
\right.
\\  \label{f}
&+& \left. 2 U^2(\K,\K_1,\K+\K_1) 
(n_{\K_1} - n_{\K+\K_1}) e^{ i (\omega_{\K} + \omega_{\K_1} 
-\omega_{\K+\K_1}) t}  
\right).
\end{eqnarray}
Proceeding in the same way one obtains
\begin{equation} \label{sigmaA}
\Sigma^A(\K,t) = 2i f(\K,t) e^{-i \omega_\K t} \theta(-t) 
\end{equation}
and
\begin{eqnarray} 
&& \Sigma^K(\K,t) =
-\frac{i  e^{-i \omega_\K t} }{V} 
\\ \nonumber
&&
\sum_{\K_1} 
\left( 2 U^2(\K,\K_1,\K-\K_1) ( 2  n_{\K_1}n_{\K-\K_1}+ n_{\K_1}+n_{\K-\K_1}+1)
e^{i ( \omega_\K - \omega_{\K_1} - \omega_{\K-\K_1}) t}\right.\nonumber\\  
 && + \left.4 U^2(\K,\K_1,\K+\K_1) 
(2n_{\K_1} n_{\K+\K_1} +n_{\K_1} +n_{\K+\K_1}  ) 
e^{i ( \omega_\K + \omega_{\K_1} - \omega_{\K+\K_1}) t}  
\right)
\label{sigmaK}
\end{eqnarray}
Let us additionally calculate $G_0^{R,A,K}$ which are needed in the main body of the text.
Inserting (\ref{G0def}) into (\ref{def2}) we get
\begin{eqnarray} \nonumber
i G_0^R(\K,t-t') =  \theta(t-t')  e^{-i\omega_\K (t-t')}
\\ \label{G0n} 
iG_0^A(\K,t-t') = - \theta(t'-t)  e^{-i\omega_\K (t-t')}
\\ \nonumber
iG_0^K(\K,t-t') =  (2 n_\K +1)   e^{-i\omega_\K (t-t')}.
\end{eqnarray}

\subsection{Derivation of the $\gamma_k$ and $\Delta_k$ coefficients}

Now we integrate $\Sigma^{R,A}$ given by (\ref{sigmaR}) and (\ref{sigmaA}) over time. We obtain
\begin{eqnarray*}
\int \mbox{d} t \, \Sigma^R(\K,t)e^{i \omega_\K t} &=& -2 i \int_0^\infty \mbox{d} t \, f(\K,t)
\\
\int \mbox{d} t \, \Sigma^A(\K,t) e^{i \omega_\K t} &=& 2 i \int_{-\infty}^0 \mbox{d} t \, f(\K,t).
\end{eqnarray*}
Inserting (\ref{f}) we obtain
\begin{eqnarray} \label{sigmaIRA}
\int \mbox{d} t \, \Sigma^R(\K,t) e^{i \omega_\K t} &=& \Delta_k - i \gamma_k
\\ \nonumber
\int \mbox{d} t \, \Sigma^A(\K,t)  e^{i \omega_\K t} &=& \Delta_k + i \gamma_k
\end{eqnarray}
where
\begin{equation} \label{delgam}
\Delta_k = 2 \int_0^\infty \mbox{d} t \, \mbox{Im}(f(\K,t))
\ \ \ \ \ \ 
\gamma_k = 2 \int_0^\infty \mbox{d} t \, \mbox{Re}(f(\K,t)).
\end{equation}
Look at the formula for $\gamma_k$ in more detail. From (\ref{f}) and (\ref{delgam}) we obtain
\begin{eqnarray} \nonumber 
\gamma_k &=&  \frac{2\pi}{V} \sum_{\K_1} \left( U^2(\K,\K_1,\K-\K_1) (n_{\K_1}+n_{\K-\K_1}+1)
 \delta ( \omega_\K -\omega_{\K_1} - \omega_{\K-\K_1})  
\right.
\\  \label{ga}
 &+& \left. 2 U^2(\K,\K_1,\K+\K_1) (n_{\K_1} - n_{\K+\K_1}) 
\delta (\omega_{\K} + \omega_{\K_1} -\omega_{\K+\K_1})  
\right).
\end{eqnarray}
To calculate the $\gamma_k$ coefficient we use the  
thermodynamic limit i.e. we  we substitute the sum by an integral 
$\frac{1}{V} \sum_{\K_1}  \rightarrow \frac{1}{(2\pi)^3} \int \mbox{d} \K_1 $.
After this substitution we obtain the formula for the $\gamma_k$ coefficient which is the same as the 
expressions given in \cite{Castin,Liu}.

Now consider the time integral over $\Sigma^K$ given by
(\ref{sigmaK}):
\begin{eqnarray} \nonumber
&& \int \mbox{d}t \, \Sigma^K(\K,t)  e^{i\omega_\K t} = - i  
\frac{1}{V} 
\\ \nonumber
&& \sum_{\K_1} 
\left( 2 U^2(\K,\K_1,\K-\K_1) ( 2  n_{\K_1}n_{\K-\K_1}+ n_{\K_1}+n_{\K-\K_1}+1) 2\pi \delta ( \omega_\K - \omega_{\K_1} - \omega_{\K-\K_1})
\right.
\\ 
&& +  \left.4 U^2(\K,\K_1,\K+\K_1) 
(2n_{\K_1} n_{\K+\K_1} +n_{\K_1} +n_{\K+\K_1}  )2\pi \delta( \omega_\K + \omega_{\K_1} - \omega_{\K+\K_1})
\right)\nonumber
\end{eqnarray}
Now we use the property of the thermal mode occupation
\begin{eqnarray*}
 (2  n_{\K_1}n_{\K-\K_1}+ n_{\K_1}+n_{\K-\K_1}+1)\delta ( \omega_\K - \omega_{\K_1} - \omega_{\K-\K_1}) 
\\
=
(2n_{\K}+1)( n_{\K_1}+n_{\K-\K_1}+1)  \delta ( \omega_\K - \omega_{\K_1} - \omega_{\K-\K_1}) 
\\
\\ 
(2n_{\K_1} n_{\K+\K_1} +n_{\K_1} +n_{\K+\K_1} )
 \delta( \omega_\K + \omega_{\K_1} - \omega_{\K+\K_1})
\\
 =  (2n_\K+1) (n_{\K_1} -n_{\K+\K_1} )\delta( \omega_\K + \omega_{\K_1} - \omega_{\K+\K_1})
\end{eqnarray*}
to obtain
\begin{equation}\label{sigmaIK}
\int \mbox{d}t \, \Sigma^K(\K,t) e^{i \omega_\K t} =   - 2 i \gamma_k (2n_\K+1).
\end{equation}

\section{Keldysh formalism in the parametric amplification case} \label{KeldyshPar}

The Hamiltonian $H_0$ given by equation (\ref{H0}) has terms $\hat b_\K \hat b_{-\K}$. 
As a consequence we deal with non-zero observables 
 $\langle \hat b_\K \hat b_{-\K} \rangle$ and 
$\langle \hat b_\K^\dagger \hat b_{-\K}^\dagger \rangle$.
So we need to define a new type of Green's function 
\begin{eqnarray*}
 iG_{12}(\K,t,t') &=& \langle T_C [ b_\K(t) b_{-\K}(t') ] \rangle
 \\ 
 i G_{21}(\K,t,t') &=& \langle T_C [ b_{-\K}^\dagger(t) b_\K^\dagger(t') ] \rangle
\end{eqnarray*}
which together with the standard one
\begin{eqnarray*}
 iG_{11}(\K,t,t') = \langle T_C [ b_\K(t) b_\K^\dagger(t') ] \rangle
\end{eqnarray*}
define our system. 
Note that here we deal with the Keldysh contour and the time ordering
operator $T_C$ on that contour.
The Dyson equation take now the matrix form 
\begin{eqnarray} \label{DysonM}
 {\bf G}(\K,t,t')  =  {\bf G}_0(\K,t,t') 
 + \int_C \mbox{d} t_1 \mbox{d} t_2 \, 
 {\bf G}_0(\K,t,t_1) {\bf \Sigma}(\K,t_1,t_2) {\bf G}(\K,t_2,t')
\end{eqnarray}
where  ${\bf G}$ and ${\bf \Sigma}$ matrices are given by equation (\ref{GM}) and (\ref{SM}) respectively.
The Dyson equation (\ref{DysonM}) involves the Keldysh contour. 
As in the previous case we move to two parts of the contour and define 
$T,<,>,\tilde T$ Green's function. For example
\begin{eqnarray*}
i {\bf G}^T(\K,t,t') &=&
\left(\begin{array}{cc}
    iG_{11}^T & iG_{12}^T \\
    iG_{21}^T & iG_{22}^T
   \end{array} \right)
(\K,t,t') 
\\
&=&
\left( \begin{array}{cc}
   \langle T [ b_\K(t) b_\K^\dagger(t') ] \rangle &   \langle T [ b_\K(t) b_{-\K}(t') ] \rangle 
   \\
   \langle T [ b_{-\K}^\dagger(t) b_\K^\dagger(t') ] \rangle    & \langle T [ b^\dagger_{-\K}(t) b_{-\K}(t') ] \rangle
   \end{array} \right). 
   \end{eqnarray*}

To 	proceed with the Keldysh method in the parametric amplification case we must show that  (\ref{rel})
is for  	${\bf G}_0$, ${\bf G}$ and ${\bf \Sigma}$  matrix functions.
It is easy to check that this is indeed the case for 	${\bf G}_0$, ${\bf G}$.
We need to show the property 
\begin{equation}\label{rel4}
 {\bf \Sigma}^T + {\bf \Sigma}^{\tilde T} = {\bf \Sigma}^> + {\bf \Sigma}^<.
\end{equation}	
In the scalar case it was shown in \cite{kamenev}.
The authors do not know if such proof is correct in the matrix case.
However, we show that it holds in second order perturbation theory in a certain approximation.
We analyze the self energy matrix.
First we concentrate on the anti-diagonal term  $\Sigma_{12}$.
In second order perturbation theory it reads
\begin{eqnarray} 
\nonumber
\Sigma_{12}(\K,t_1,t_2) = \frac{i}{V} \sum_{\K_1}\left( 2 U^2(\K,\K_1,\K-\K_1) G_{0,12}(\K_1,t_1,t_2) G_{0,12}(\K-\K_1,t_1,t_2) \right. 
\\  \label{sigma12}
+ \left. 4U^2(\K,\K_1,\K+\K_1) 
G_{0,12}(\K+\K_1,t_1,t_2)G_{0,21}(-\K_1,t_1,t_2) \right)
\end{eqnarray}
It contains anti-diagonal Green's functions $G_{0,12}$ and $G_{0,21}$. 
But according to the assumption stated in Section \ref{SPar} the annihilation operators
of all the modes, while contributing to the self energy functions, undergo
evolution $\hat b_\K(t) = \hat b_\K(0) e^{-i\omega_\K t} $.
For such evolution $G_{0,12}(\K,t) = G_{0,21}(\K,t) = 0$ and as a consequence
$\Sigma_{12} = 0$. The same reasoning applies to $\Sigma_{21}$.
Now we move to $\Sigma_{11}$. One can show that in the second order perturbation it is given by 
\begin{eqnarray}
\nonumber
\Sigma_{11}(\K,t_1,t_2) = \frac{i}{V} \sum_{\K_1} 
\left( 2 U^2(\K,\K_1,\K-\K_1) G_{0,11}(\K_1,t_1,t_2)G_{0,11}(\K-\K_1,t_1,t_2) \right.
\\  \label{sigma11}
+ \left.4 
 U^2(\K,\K_1,\K+\K_1) 
G_{0,11}(\K+\K_1,t_1,t_2)G_{0,11}(\K_1,t_2,t_1) \right)
\end{eqnarray}
and has exactly the same form as $\Sigma(\K,t_1,t_2)$ given by equation (\ref{sigmaC}).
The evolution of mode operators is the same as in the scalar case 
which means that $G_{0,11}$  above is equal to $G_0$ appearing in (\ref{sigmaC}).
As a result we simply
have $\Sigma_{11}(\K,t_1,t_2) = \Sigma(\K,t_1,t_2)$.
Then (\ref{rel}) gives
 \begin{equation}\label{s11}
  \Sigma_{11}^T +  \Sigma_{11}^{\tilde T} =  \Sigma_{11}^> +  \Sigma_{11}^<.
 \end{equation}
From relation $\Sigma_{22}(\K,t,t') = \Sigma_{11}(-\K,t',t)$ we obtain 
 \begin{equation}
 \Sigma_{22}^{T,\tilde T,>,<} (\K,t,t') = \Sigma_{11}^{T,\tilde T,<,>}(-\K,t',t),
\end{equation}
 which together with (\ref{s11}) gives
 \begin{eqnarray*}
  \Sigma_{22}^T +  \Sigma_{22}^{\tilde T} =  \Sigma_{22}^> +  \Sigma_{22}^<.
 \end{eqnarray*}
As a result we find that the property given by equation (\ref{rel4}) holds.
Proceeding in the same way as in the scalar case we define $R$, $A$ and $K$ components of
${\bf G}_0$, ${\bf G}$ and ${\bf \Sigma}$  matrix function using (\ref{def2}).
In the scalar case the Dyson equations (\ref{Dyson1}) and (\ref{Dyson3})
are derived from (\ref{Dyson}) by changing the Keldysh time contour into a single
time axis. As this change deals only with time arguments, the Dyson equations in the matrix
case shall be the same as in the scalar case i.e.
\begin{eqnarray} \label{Dyson1M}
&& {\bf G}^{R,A} = {\bf G}_0^{R,A} + {\bf G}_0^{R,A} {\bf \Sigma}^{R,A} {\bf G}^{R,A} 
\\ \label{Dyson3M}
&& {\bf G}^K = (1+{\bf G}^R{\bf \Sigma}^R){\bf G}_0^K(1+{\bf \Sigma}^A {\bf G}^A) + {\bf G}^R{\bf \Sigma}^K{\bf G}^A.
\end{eqnarray}

We now derive the ${\bf \Sigma}^R= {\bf \Sigma}^T  -{\bf \Sigma}^< $ matrix. 
Using the definition of $T$ and $<$ we obtain
\begin{eqnarray*}
 {\bf \Sigma}^T(\K,t_1,t_2) &=& 
\left(\begin{array}{cc}
    \Sigma_{11}^T(\K,t_1,t_2) & 0 \\
    0 & \Sigma_{22}^T(\K,t_1,t_2)
   \end{array} \right) 
\\
 &=&
\left(\begin{array}{cc}
    \Sigma_{11}^T(\K,t_1,t_2) & 0 \\
    0 & \Sigma_{11}^T(-\K,t_2,t_1)
   \end{array} \right) 
\end{eqnarray*}
and
\begin{eqnarray*}
 {\bf \Sigma}^<(\K,t_1,t_2) &=& 
\left(\begin{array}{cc}
    \Sigma_{11}^<(\K,t_1,t_2) & 0 \\
    0 & \Sigma_{22}^<(\K,t_1,t_2)
   \end{array} \right) 
\\ 
  &=&
\left(\begin{array}{cc}
    \Sigma_{11}^<(\K,t_1,t_2) & 0 \\
    0 & \Sigma_{11}^>(-\K,t_2,t_1)
   \end{array} \right). 
\end{eqnarray*}
Note that in the above $ \Sigma_{22}^<(\K,t_1,t_2) = \Sigma_{11}^>(-\K,t_2,t_1) $.
As a result we obtain
\[ {\bf \Sigma}^R(\K,t_1,t_2) = 
\left(\begin{array}{cc}
    \Sigma_{11}^R(\K,t_1,t_2) & 0 \\
    0 & \Sigma_{11}^A(-\K,t_2,t_1)
   \end{array} \right). 
\]
Proceeding in the same way we obtain
\[ {\bf \Sigma}^A(\K,t_1,t_2) = 
\left(\begin{array}{cc}
    \Sigma_{11}^A(\K,t_1,t_2) & 0 \\
    0 & \Sigma_{11}^R(-\K,t_2,t_1)
   \end{array} \right) 
\]
and
\[ {\bf \Sigma}^K(\K,t_1,t_2) = 
\left(\begin{array}{cc}
    \Sigma_{11}^K(\K,t_1,t_2) & 0 \\
    0 & \Sigma_{11}^K(-\K,t_2,t_1)
   \end{array} \right). 
\]
The same applies to the ${\bf G}^{R,A,K}$ function which reads
\begin{eqnarray*}
 {\bf G}^{R,A,K}(\K,t_1,t_2) 
&=& \left(\begin{array}{cc}
    G_{11}^{R,A,K}(\K,t_1,t_2) & G_{12}^{R,A,K}(\K,t_1,t_2)   \\
    G_{21}^{R,A,K}(\K,t_1,t_2) & G_{22}^{R,A,K}(\K,t_1,t_2)   
   \end{array} \right) 
\\
&=& \left(\begin{array}{cc}
    G_{11}^{R,A,K}(\K,t_1,t_2) & G_{12}^{R,A,K}(\K,t_1,t_2)   \\
    G_{21}^{R,A,K}(\K,t_1,t_2) & G_{11}^{A,R,K}(-\K,t_2,t_1)   
   \end{array} \right)
\end{eqnarray*}
Note that in the above 
$ G_{22}^{R,A,K}(\K,t_1,t_2) =  G_{11}^{A,R,K}(-\K,t_2,t_1) $ which comes directly from the definitions.
Now we analyse approximation (\ref{sap}) in matrix case.
It takes the form
\begin{eqnarray}\nonumber
&& \left(\begin{array}{cc}
    \Sigma_{11}^{R,A,K}(\K,t) e^{i\omega_\K t} & 0 \\
    0 & \Sigma_{22}^{R,A,K}(\K,t) e^{-i\omega_\K t}
   \end{array} \right)
\\ \label{sap2}
&& = \delta (t)
    \int \mbox{d} \tau \, 
\left(\begin{array}{cc}
    \Sigma_{11}^{R,A,K}(\K,\tau)  e^{i\omega_\K \tau}  & 0 \\
    0 & \Sigma_{11}^{A,R,K}(-\K,-\tau) e^{-i\omega_\K \tau}
   \end{array} \right). 
\end{eqnarray}
Let us now use the fact that  $\Sigma_{11}$ given by equation (\ref{sigma11}) is equal to
$\Sigma$ given by equation (\ref{sigmaC}) as we discussed above.
Then $\Sigma_{11}^{R,A,K} = \Sigma^{R,A,K}$ given by
equations (\ref{sigmaR}), (\ref{sigmaA}) and (\ref{sigmaK}). Thus the time integrals present in (\ref{sap2}) 
can be directly connected to integrals in (\ref{sigmaIRA}) and (\ref{sigmaIK}) leading to
\begin{eqnarray*}
 \left(\begin{array}{cc}
    \Sigma_{11}^{R}(\K,t) e^{i\omega_\K t} & 0 \\
    0 & \Sigma_{22}^{R}(\K,t) e^{-i\omega_\K t}
   \end{array} \right)
   = \delta(t) 
\left(\begin{array}{cc}
    \Delta_k-i\gamma_k & 0 \\
    0 & \Delta_k + i \gamma_k
   \end{array} \right), 
\end{eqnarray*}
\begin{eqnarray*}
 \left(\begin{array}{cc}
    \Sigma_{11}^A(\K,t) e^{i\omega_\K t} & 0 \\
    0 & \Sigma_{22}^A(\K,t) e^{-i\omega_\K t}
   \end{array} \right)
   = \delta(t) 
\left(\begin{array}{cc}
    \Delta_k+i\gamma_k & 0 \\
    0 & \Delta_k - i \gamma_k
   \end{array} \right),
\end{eqnarray*}
and
\begin{eqnarray*}
 \left(\begin{array}{cc}
    \Sigma_{11}^K(\K,t) e^{i\omega_\K t} & 0 \\
    0 & \Sigma_{22}^K(\K,t) e^{-i\omega_\K t}
   \end{array} \right)
   = \delta(t) (-2i) \gamma_k(n_\K+1)
\left(\begin{array}{cc}
    1 & 0 \\
    0 & 1
   \end{array} \right).
\end{eqnarray*}

\subsection{Green's function $G_0$}

The evolution of annihilation operators due to Hamiltonian (\ref{H0}) is given by (\ref{bt}).
We rewrite it as
\begin{eqnarray*}
 \hat b_\K(t) = A(\Delta,t) \hat b_\K(0) e^{-i\omega t} + B(\Delta,t) \hat b_{-\K}^\dagger(0) e^{-i\omega t}
\end{eqnarray*}
where 
\begin{eqnarray*}
 A(\Delta,t) = \cosh \Omega t + i \frac{\Delta}{\Omega} \sinh \Omega t
\ \ \ \ \ \ 
B(\Delta,t) = -i \frac{\delta}{\Omega} \sinh\Omega t
 \end{eqnarray*}
 and $\Delta =  \omega - \omega_k$, $\Omega = \sqrt{\delta^2 - \Delta^2}$.
The matrix Green's function  ${\bf G}_0$ corresponding to that evolution is given by 
 \begin{eqnarray*}
  i{\bf G}_0^R(\K,t,t') = 
  \left(\begin{array}{cc}
    A_R(\Delta,t,t')   & B_R(\Delta,t,t')    \\
      -B_R^*(\Delta,t,t')  &  -A_R(\Delta,t',t) 
   \end{array} \right) \theta(t-t') 
 \end{eqnarray*}
\begin{eqnarray*}
  i{\bf G}_0^A(\K,t,t') = 
  \left(\begin{array}{cc}
    -A_R(\Delta,t,t')   & -B_R(\Delta,t,t')    \\
      B_R^*(\Delta,t,t')  &  A_R(\Delta,t',t) 
   \end{array} \right) \theta(t'-t) 
 \end{eqnarray*}
where
\begin{eqnarray*}
 A_R(\Delta,t,t') =\\
 A(\Delta,t-t')e^{-i\omega (t-t')}\theta(t)\theta(t')+A(\Delta,t) e^{-i\omega t + i (\omega-\Delta) t'} \theta(t)\theta(-t') \\
 + A^*(\Delta,t')\theta(t')\theta(-t) e^{i\omega t' - i (\omega-\Delta) t}
+e^{-i(\omega - \Delta) (t-t')}\theta(-t)\theta(-t') 
\\
\\
 B_R(\Delta,t,t') =\\
 - B(\Delta,t-t')e^{-i\omega (t+t')}\theta(t)\theta(t')
 - B(\Delta,t)
 e^{-i(\omega t + (\Omega-\Delta) t')}\theta(t)\theta(-t')\\
 + B(t')\theta(t')\theta(-t)
e^{-i(\omega t' + (\omega-\Delta) t)}
\end{eqnarray*}
Additionally 
\begin{eqnarray*}
  i{\bf G}_0^K(\K,t,t') = 
  \left(\begin{array}{cc}
    A_K(\Delta,t,t')   & B_K(\Delta,t,t')    \\
      B_K^*(\Delta,t,t')  &  A_K(\Delta,t',t) 
   \end{array} \right)  
 \end{eqnarray*}
where
\begin{eqnarray*}
A_K(\Delta,t,t') =(2n_\K+1)\\
 \left(\left( A(\Delta,t)A^*(\Delta,t')
+B(\Delta,t)B^*(\Delta,t') \right) e^{-i\omega(t-t')} \theta(t)\theta(t')\right.\\
+\left. A(\Delta,t) e^{-i(\omega t - (\omega - \Delta) t') } \theta(t)\theta(-t')+A^*(\Delta,t') e^{-i((\omega - \Delta) t - \omega t')} \theta(-t)\theta(t')\right.\\
+\left. e^{-i(\omega - \Delta) (t -t') } \theta(-t)\theta(-t') \right)
\\
\\
B_K(\Delta,t,t') =  (2n_\K+1)\\
 \left( \left( A(\Delta,t)B(\Delta,t') 
+ B(\Delta,t)A(\Delta,t') \right) e^{-i\omega(t+t')} \theta(t)\theta(t')
\right.
\\ 
+   \left.  B(\Delta,t) e^{-i(\omega t + (\omega - \Delta) t')} \theta(t) \theta(-t')
+ B(\Delta,t')  e^{-i((\omega - \Delta) t + \omega t')} \theta(-t) \theta(t') \right).
\end{eqnarray*}

\subsection{Number squeezing parameter}\label{nsp}

Below we derive the number squeezing parameter given in Section \ref{SPar}.
In the system we have the symmetry of the state with respect to the change $\K \rightarrow -\K$ which implies that
$ n_\K(t)= n_{-\K}(t) $ and
\begin{eqnarray*}
\langle \left( \hat b_\K^\dagger \hat b_\K - \hat b_{-\K}^\dagger \hat b_{-\K} \right)^2 \rangle
&=& \langle \hat b_\K^\dagger \hat b_\K \hat b_\K^\dagger \hat b_\K  + 
 \hat b_{-\K}^\dagger \hat b_{-\K} \hat b_{-\K}^\dagger \hat b_{-\K} \rangle-\langle \hat b_\K^\dagger \hat b_\K  \hat b_{-\K}^\dagger \hat b_{-\K}
 + \hat b_{-\K}^\dagger \hat b_{-\K}  \hat b_\K^\dagger \hat b_\K \rangle 
\\
 &=& 2 \left( \langle \hat b_\K^\dagger \hat b_\K \hat b_\K^\dagger \hat b_\K \rangle
  -  \langle  \hat b_\K^\dagger \hat b_\K  \hat b_{-\K}^\dagger \hat b_{-\K} \rangle \right).
\end{eqnarray*}
In the lowest order of approximation when using the Wick theorem we have
\begin{eqnarray*}
 \langle \hat b_\K^\dagger \hat b_\K \hat b_\K^\dagger \hat b_\K \rangle &\simeq& n_\K(t)(2 n_\K(t)+1)
 \\
 \langle  \hat b_\K^\dagger \hat b_\K  \hat b_{-\K}^\dagger \hat b_{-\K} \rangle &\simeq& n_\K(t)n_{-\K}(t) 
 + \langle \hat b_\K \hat b_{-\K} \rangle \langle \hat b_\K^\dagger \hat b_{-\K}^\dagger \rangle.
\end{eqnarray*}
As $\hat b_\K(t)$ commutes with  $\hat b_{-\K}(t)$ we have
$\langle \hat b_\K(t) \hat b_{-\K}(t) \rangle = iG_{12}^>(\K,t,t) = iG_{12}^<(\K,t,t) $
Since $G_{12}^K =G_{12}^> + G_{12}^< $ we get
\begin{eqnarray*} 
 \langle \hat b_\K(t) \hat b_{-\K}(t) \rangle  = \frac{1}{2} i G_{12}^K(\K,t,t).
\end{eqnarray*}
Proceeding the same way we obtain
\begin{eqnarray*}
 \langle \hat b_\K^\dagger(t) \hat b_{-\K}^\dagger(t) \rangle  = \frac{1}{2} i G_{21}^K(\K,t,t).
\end{eqnarray*}
As a result the numerator of the $s$ parameter takes the form
\begin{eqnarray*}
\langle \Delta \hat{n}_\K^2(t) \rangle &=&
2 \left( n_\K(t)(n_\K(t)+1) - \frac{1}{4}\left(iG_{12}^K(\K,t,t)\right)\left(iG_{21}^K(\K,t,t)\right) \right) 
\\
&=&
\frac{1}{2} \left(  
(2n_\K+1)^2 \frac{ 2e^{-2\gamma_k t} \gamma_k\delta \sinh(2\delta t) + e^{-4 \gamma_k t} \delta^2 - \gamma_k^2 }{
\delta^2 -\gamma_k^2 } - 1 
\right)
\end{eqnarray*}
where we have used equations (\ref{GPK}), (\ref{ab}) and (\ref{nkt}).

\subsection{Justification of the Wick decomposition}\label{Wick}

In Section \ref{SPar} the following quantities we needed 
\begin{eqnarray*}
 \langle \hat b_\K^\dagger(t) \hat b_\K(t) \hat b_\K^\dagger(t) \hat b_\K(t) \rangle 
 \ \ \ \ \ \ 
 \langle  \hat b_\K^\dagger(t) \hat b_\K(t)  \hat b_{-\K}^\dagger(t) \hat b_{-\K}(t) \rangle. 
\end{eqnarray*}
The above can be rewritten as
\begin{eqnarray*}
 \langle \hat b_\K^\dagger(t) \hat b_\K^\dagger(t) \hat b_\K(t) \hat b_\K(t) \rangle +
 \langle  \hat b_\K^\dagger(t) \hat b_\K(t) \rangle
 \ \ \ \ \ \  
 \langle  \hat b_\K^\dagger(t) \hat b_{-\K}^\dagger(t)\hat b_\K(t)   \hat b_{-\K}(t) \rangle. 
\end{eqnarray*}
The way to calculate them in a perturbative manner via quantum field theory methods is described in \cite{gorkov}.
Here we shall describe it briefly. We proceed in the same way as when calculating one body observables such
as Green's function. So we take in the above quantities different times and apply the time ordering operator i.e.
\begin{displaymath}
 \langle T[ \hat b_\K(t_3) \hat b_\K(t_4)  b_\K^\dagger(t_1)  \hat b_\K^\dagger(t_2)]  \rangle 
\end{displaymath}
We proceed in the same way as when calculating the one body Green's function i.e. expanding the above into Feynmann
diagrams. Unlike the diagrams for the one body Green's function, all these diagrams have four external lines: two
incoming which we call $1$ and $2$ and two out-coming $3$ and $4$.
While calculating the above observable we need to consider only connected diagrams  as in the one body case.
All the connected diagrams can be divided into two groups. First group 
contains the diagrams in which points $1$ and $3$ as well as points $2$ and $4$, 
are connected by a  sequence of pairings, while the points $1$ and $2$, $1$ and $4$, $2$ and $3$,  $3$ and $4$ 
are isolated from each other.
Such diagrams decompose into two separate parts which are not connected to by any lines.
Moreover, we assign to the same group all diagrams in which $1$ is connected to $4$ and $2$ to  $3$, while
$1$ and $2$, $1$ and $3$, $3$ and $4$, $2$ and $4$ are not connected.
It is not hard to see that all such diagrams give
\begin{eqnarray*}
 G(\K,t_3,t_1) G(\K,t_4,t_2) + G(\K,t_4,t_1) G(\K,t_3,t_2)  
\end{eqnarray*}
This part gives the Wick theorem decomposition
\begin{eqnarray*}
&& \langle \hat b_\K^\dagger(t) \hat b_\K^\dagger(t) \hat b_\K(t) \hat b_\K(t) \rangle \rightarrow
2 \langle \hat b_\K^\dagger(t) \hat b_\K(t) \rangle^2
 \\ 
&& \langle  \hat b_\K^\dagger(t) \hat b_{-\K}^\dagger(t)\hat b_\K(t)   \hat b_{-\K}(t) \rangle
\rightarrow  \langle \hat b_\K^\dagger(t) \hat b_\K(t) \rangle^2 + |\langle \hat b_\K(t)   \hat b_{-\K}(t) \rangle|^2
\end{eqnarray*}
The other group of diagrams consists of the set of diagrams which do not decompose into separate parts.
In the case of Hamiltonian (\ref{Hint}) such diagrams are proportional to $1/V$.

\section*{References}

\providecommand{\newblock}{}

\end{document}